\documentstyle[preprint,pre,aps]{revtex}
\tightenlines
\begin{document}
\draft
\title{Renormalization analysis of intermittency in two coupled maps}
\author{
        Sang-Yoon Kim
        }
\address{
Department of Physics \\
Kangwon National University\\
Chunchon, Kangwon-Do 200-701, Korea
}
\maketitle

\begin{abstract}
The critical behavior for intermittency is studied in two coupled
one-dimensional (1D) maps. We find two fixed maps of an approximate
renormalization operator in the space of coupled maps. Each fixed 
map has a common relavant eigenvaule associated with the scaling
of the control parameter of the uncoupled one-dimensional map.
However, the relevant ``coupling eigenvalue'' associated with
coupling perturbation varies depending on the fixed maps.
These renormalization results are also confirmed for a
linearly-coupled case.

\end{abstract}

\pacs{PACS numbers : 05.45.+b, 03.20.+i, 05.70.Jk\\
      Keywords: Renormalization, Intermittency, Coupled Maps}

%
%

\narrowtext

A route to chaos via intermittency in the one-dimensional (1D) map is
associated with a saddle-node bifurcation \cite{MP}. Intermittency 
just preceding a saddle-node bifurcation to a periodic attractor is 
characterized by the occurrence of intermittent alternations between 
regular behavior and chaotic behavior. Scaling relations for the 
average duration of regular behavior in the presence of noise have 
been first established \cite{EH} by considering a Langevin equation 
describing the map near the intermittency threshold and using 
Fokker-Plank techniques. The same scaling results for intermittency 
have been later found \cite{HH} by employing the same 
renormalization-group equation \cite{Feigenbaum} for period doubling 
with a mere change of boundary conditions appropriate to a 
saddle-node bifurcation.

Recently, universal scaling results of period doubling for the 1D map 
have been generalized to the coupled 1D maps 
\cite{Kapral,Kuznet,Aranson,Kim1,Kim2}, which are used to simulate 
spatially extended systems with effectively many degrees of freedom 
\cite{Kaneko}. It has been found that the critical scaling behaviors 
of period doubling for the coupled 1D maps are much richer than those 
for the uncoupled 1D map \cite{Kim1,Kim2}. These results for the 
abstract system of the coupled 1D maps are also confirmed in the real 
system of the coupled oscillators \cite{Kim3}. Similarly, the scaling 
results of the higher period $p$-tuplings $(p=,3,4,...)$ in the 1D map 
are also generalized to the coupled 1D maps \cite{Kim4}. Here we are 
interested in another route to chaos via intermittency in coupled 1D 
maps. Using a renormalization method, we extend the scaling results of 
intermittency for the 1D map to two coupled 1D maps.

Consider a map $T$ consisting of two identical 1D maps coupled
symmetrically, 
\begin{equation}
T: \left \{ 
\begin{array}{l}
x_{n+1}=f(x_n)+g(x_n,y_n), \\ 
y_{n+1}=f(y_n)+g(y_n,x_n),
\end{array}
\right.  \label{eq:TCM}
\end{equation}
where the subscript $n$ denotes a discrete time, $f(x)$ is a 1D map 
with a quadratic maximum, and $g(x,y)$ is a coupling function 
obeying a condition, 
\begin{equation}
g(x,x)=0\;\;{\rm for\;\;any}\;\;x.  \label{eq:CC}
\end{equation}

The two-coupled 1D map (\ref{eq:TCM}) is called a symmetric map 
because it has an exchange symmetry such that 
\begin{equation}
{\sigma}^{-1}T{\sigma}({\bf z})=T({\bf z})\;\;
{\rm for\;\;all\;\;}{\bf z},
\label{eq:ES}
\end{equation}
where ${\bf z}=(x,y)$, $\sigma$ is an exchange operator acting on 
${\bf z}$ such that $\sigma {\bf z}=(y,x)$, and ${\sigma}^{-1}$ is 
its inverse. The set of all fixed points of $\sigma$ forms a 
synchronization line $y=x$ in the state space. It follows from 
Eq.~(\ref{eq:ES}) that the exchange operator $\sigma$ commutes with 
the symmetric map $T$, i.e., $\sigma T = T \sigma$. Thus the 
synchronization line becomes invariant under $T$. An orbit is called 
a(n) (in-phase) synchronous orbit if it lies on the invariant 
synchronization line, i.e., it satisfies 
\begin{equation}
x_n=y_n\;\;{\rm for\;\;all\;\;}n.  \label{IO}
\end{equation}
Otherwise, it is called an (out-of-phase) asynchronous orbit.

Let us introduce new coordinates $X$ and $Y$, 
\begin{equation}
X={\frac{{x+y} }{2}},\;\;\;Y={\frac{{x-y} }{2}}.  \label{eq:NC}
\end{equation}
Then the map (\ref{eq:TCM}) becomes 
\begin{eqnarray}
X_{n+1} &=& F(X_n,Y_n)  \nonumber \\
&=& {\frac{1 }{2}}\; [f(X_n+Y_n)+f(X_n-Y_n)]  \nonumber \\
&&+{\frac{1 }{2}}\; [g(X_n+Y_n,X_n-Y_n)+g(X_n-Y_n,X_n+Y_n)],  
\nonumber \\
&& \label{eq:NTCM}          \\
Y_{n+1} &=& G(X_n,Y_n)  \nonumber \\
&=& {\frac{1 }{2}}\; [f(X_n+Y_n)-f(X_n-Y_n)]  \nonumber \\
&&+{\frac{1 }{2}}\; [g(X_n+Y_n,X_n-Y_n)-g(X_n-Y_n,X_n+Y_n)]. 
 \nonumber
\end{eqnarray}
This map is invariant under the reflection $Y \rightarrow -Y$, and 
hence the invariant synchronization line becomes $Y=0$. Then the 
synchronous orbit of the old map (\ref{eq:TCM}) becomes the orbit of 
this new map with $Y=0$. Furthermore, the $X$-coordinate of the 
synchronous orbit satisfies the uncoupled 1D map, i.e., 
$X_{n+1}=f(X_n)$, because the coupling function $g$ obeys the
condition (\ref{eq:CC}).

Stability of a synchronous orbit of period $p$ is determined from the 
Jacobian matrix $M$ of $T^p$, which is given by the $p$ product of 
the linearized map $DT$ of the map (\ref{eq:NTCM}) along the orbit 
\begin{eqnarray}
M &=& {\prod_{n=1}^{p}} DT(X_n,0)  \nonumber \\
&=& {\prod_{n=1}^{p}} \left ( 
\begin{array}{cc}
f^{\prime}(X_n) & 0 \\ 
0 & f^{\prime}(X_n)-2G(X_n)
\end{array}
\right ),  \label{eq:JM}
\end{eqnarray}
where $f^{\prime}(X)=df(X)/dX$ and $G(X)= \partial g(X,Y)/ \partial 
Y |_{Y=X}$. The eigenvalues of $M$, called the Floquet (stability) 
multipliers of the orbit, are 
\begin{equation}
\lambda_1 = {\prod_{n}^{p}} f^{\prime}(X_n),\;\; \lambda_2 = 
{\prod_{n}^{p}} [f^{\prime}(X_n) - 2G(X_n)].  
\label{eq:SM}
\end{equation}
Note that $\lambda_1$ is just the Floquet multiplier for the case of 
the uncoupled 1D map and the coupling affects only $\lambda_2$.

Consider a synchronous saddle-node bifurcation to a synchronous 
periodic orbit. The synchronous periodic orbit is stable when both 
Floquet multipliers lie inside the unit circle, i.e., 
$|\lambda_j| < 1$ for $j=1$ and $2$. Thus its stable region in the 
parameter plane is bounded by four bifurcation lines, i.e., those 
curves determined by the equations $\lambda_j=\pm1$ $(j=1,2)$.  
When a Floquet multiplier $\lambda_j$ increases thorugh 
$1$, the stable synchronous periodic orbit loses its stability 
via saddle-node or pitchfork bifurcation. On the other hand,
when a Floquet multiplier $\lambda_j$ decreases thorugh $-1$, it 
becomes unstable via period-doubling bifurcation. (For more details
on bifurcations, refer to Ref.~\cite{Guckenheimer}.)

Here we are interested in intermittency just preceding the synchronous
saddle-node bifurcation. Employing an approximate renormalization 
operator \cite{Kim2,Greene,Mao,Lahiri} which includes a truncation, we 
generalize the 1D scaling results for intermittency to the case of two 
coupled 1D maps. We thus find two fixed maps of the approximate 
renormalization operator. They have a common relavant eigenvaule 
associated with the scaling of the control parameter of the uncoupled 
1D map. However, the relevant ``coupling eigenvalue'' associated with
coupling perturbation varies depending on the fixed maps. 

Truncating the map (\ref{eq:NTCM}) at its quadratic terms, we have 
\begin{equation}
T_{{\bf P}}: \left \{ 
\begin{array}{l}
X_{n+1}=A+BX_n + C X_n^2 + F Y_n^2 \\ 
Y_{n+1}=D Y_n + E X_n Y_n
\end{array}
\right. ,  \label{eq:TM}
\end{equation}
which is a six-parameter family of coupled maps. Other terms do not 
appear because $F(X,Y)$ and $G(X,Y)$ in Eq.~(\ref{eq:NTCM}) are even 
and odd in $Y$, respectively. Here ${\bf P}$ represents the six 
parameters, i.e., ${\bf P} = (A,B,C,D,E,F)$. The construction of 
Eq.~(\ref{eq:TM}) corresponds to a truncation of the infinite 
dimensional space of coupled maps to a six-dimensional space. The 
parameters $A,\;B,\;C,\;D,\;E,\;$ and $F$ can be regarded as the 
coordinates of the truncated space.

We look for fixed points of the renormalization operator ${\cal R}$ in
the truncated six-dimensional space of coupled maps, 
\begin{equation}
{\cal R} (T) = \Lambda T^2 \Lambda^{-1}.  \label{eq:RO}
\end{equation}
Here the rescaling operator $\Lambda$ is given by 
\begin{equation}
\Lambda = \left ( 
\begin{array}{cc}
\alpha & 0 \\ 
0 & \alpha
\end{array}
\right ),  \label{eq:SO}
\end{equation}
where $\alpha$ is a rescaling factor.

The operation ${\cal R}$ in the truncated space can be represented by 
a transformation of parameters, i.e., a map from ${\bf P} \equiv 
(A,B,C,D,E,F)$ to ${\bf P^{\prime}} \equiv
(A^{\prime},B^{\prime},C^{\prime},D^{\prime},E^{\prime},F^{\prime}),$ 
\begin{mathletters}
\label{eq:PT}
\begin{eqnarray}
A^{\prime}&=& \alpha A (1+B+AC),  \label{eq:RGTA} \\
B^{\prime}&=& B (B+2AC),  \label{eq:RGTB} \\
C^{\prime}&=& {\frac{C }{\alpha}} (B+B^2+2AC),  \label{eq:RGTC} \\
D^{\prime}&=& D (D+AE),  \label{eq:RGTD} \\
E^{\prime}&=& {\frac{E }{\alpha}} (BD+D+AE),  \label{eq:RGTE} \\
F^{\prime}&=& {\frac{F }{\alpha}} (2AC+B+D^2) .  \label{eq:RGTF}
\end{eqnarray}
The fixed point ${\bf P}^* =(A^*,B^*,C^*,D^*,E^*,F^*)$ of this map 
can be determined by solving ${\bf P}^{\prime}={\bf P}$. We thus find 
two solutions associated with a saddle-node bifurcation, as will be 
seen below. The map (\ref {eq:TM}) with a solution ${\bf P^*}$ 
$(T_{{\bf P^*}})$ is the fixed map of the renormalization 
transformation ${\cal R}$; for brevity $T_{{\bf P^*}}$ will be denoted 
as $T^*$.

For a saddle-node bifurcation at $x=0$, the 1D map $f(x)$ satisfies 
\end{mathletters}
\begin{equation}
f(0)=0,\;\;\;f^{\prime}(0)=1.  \label{eq:bc}
\end{equation}
Hence the function $F(X,Y)$ in Eq.~(\ref{eq:NTCM}) obeys 
\begin{equation}
\left. F(0,0)=0,\;\;\;{\frac{{\partial F} }{{\partial X}}} 
\right|_{(0,0)}=1.  
\label{eq:NBC}
\end{equation}

We first note that Eqs.~(\ref{eq:RGTA})-(\ref{eq:RGTC}) are only for 
$A,\;B,\;C,$ and $\alpha$. We find one solution for $A^*,\;B^*,\;C^*,$ 
and $\alpha$ satisfying the conditions (\ref{eq:NBC}), 
\begin{equation}
\alpha=2,\;\;A^*=0,\;\;B^*=1,\;\;C^*:{\rm arbitrary\;number}.
\end{equation}
Substituting the values of $A^*,\;B^*$ and $\alpha$ into 
Eqs.~(\ref{eq:RGTD})-(\ref{eq:RGTF}), we have two solutions for $D^*,
\;E^*,$ and $F^*$, 
\begin{mathletters}
\label{eq:FP}
\begin{eqnarray}
D^*&=&0,\;\;E^*=0,\;\;F^*=0, \\
D^*&=&1,\;\;E^*:{\rm arbitrary\;number},\;\;F^*:
{\rm arbitrary\;number}.
\end{eqnarray}
\end{mathletters}
These two solutions are associated with intermittency in the coupled 
1D maps, as will be seen below. Hereafter we will call each map from 
the top as the $I$ and $E$ map, respectively, as listed in Table 
\ref{FP}.

Consider an infinitesimal perturbation $\epsilon \, \delta {\bf P}$ 
to a fixed point ${\bf P}^*$ of the transformation of parameters 
(\ref{eq:RGTA})-(\ref{eq:RGTF}). Linearizing the transformation at 
${\bf P}^*$, we obtain the equation for the evolution of $\delta 
{\bf P}$, 
\begin{equation}
\delta {\bf P}^{\prime}= J \delta {\bf P},
\end{equation}
where $J$ is the Jacobian matrix of the transformation at ${\bf P}^*$.

Since the $6 \times 6$ Jacobian matrix $J$ decomposes into smaller 
blocks, one can easily obtain its eigenvalues. Two of them are 
\begin{equation}
\left. \lambda_1 = {\frac{{\partial C^{\prime}} }{{\partial C}}} 
\right|_{{\bf P^*}} =1, \;\;\; \left. \lambda_2 = {\frac{{\partial 
F^{\prime}} }{{\partial F}}} \right|_{{\bf P^*}} = {\frac{{1+D^{*2}}}
{2}}.
\end{equation}
Here $\lambda_1$ is an eigenvalue associated with scale change in $X$, 
and hence $C^*$ is arbitrary. The eigenvalue $\lambda_2$ is also 
associated with scale change in $Y$ in the case $D^*=1$; this case 
corresponds to the $E$ map. Thus $F^*$ for this case becomes 
arbitrary. However, in the case $D^*=0$ corresponding to the $I$ map, 
$\lambda_2$ becomes an irrelevant eigenvalue. Note that the $I$ map 
is invariant under a scale change in $Y$ because $F^*=0$.

The remaining four eigenvalues are those of the following $2 \times 2$
blocks, 
\begin{eqnarray}
\left. M_1 = {\frac{ {\partial (A^{\prime},B^{\prime})} }
{{\partial (A,B)} }} \right|_{{\bf P^*}} = \left( 
\begin{array}{cc}
4 & 0 \\ 
2\,C^* & 2
\end{array}
\right), \\
\left. M_2 = {\frac{ {\partial (D^{\prime},E^{\prime})} }
{{\partial (D,E)} }} \right|_{{\bf P^*}} = \left( 
\begin{array}{cc}
2\,D^* & 0 \\ 
E^* & D^*
\end{array}
\right).
\end{eqnarray}
The two eigenvalues of $M_i$ $(i=1,2)$ are called $\delta_i$ and 
$\delta_i^{\prime}$, and they are listed in Table \ref{EV}.

The two $I$ and $E$ maps have common eigenvalues of $M_1$. They are $
\delta_1=4$ and $\delta_1^{\prime}=2$, which are just the relevant 
eigenvalues \cite{HH} for the case of uncoupled 1D maps. Here the 
largest relevant eigenvalue $\delta_1$ is associated with scaling of 
the control parameter of the 1D map near the intermittency threshold.

The eigenvalues $\delta_2$ and $\delta_2^{\prime}$ of $M_2$ are 
associated with coupling perturbations. These eigenvalues will be 
referred to as ``coupling eigenvalues'' (CE's). The submatrix $M_2$ 
for the $I$ map becomes a null matrix, and hence there exist no CE's. 
On the other hand, the $E$ map has a relevant CE $\delta_2=2$ and a 
marginal CE $\delta_2^{\prime}=1$. Here the relevant CE $\delta_2$ 
is associated with scaling of the coupling parameter, while the 
marginal one $\delta_2^{\prime}$ is associated with the arbitrary 
constant $E^*$.

We also obtain the Floquet multipliers $\lambda_1^*$ and 
$\lambda_2^*$ of the fixed point $(0,0)$ of the fixed map $T^*$ of 
the renormalization transformation ${\cal R}$. They are given by 
\begin{equation}
\lambda_1^*=1,\;\;\;\lambda_2^*=D^*.  
\label{eq:CSM}
\end{equation}
The $I$ and $E$ maps have a common Floquet multiplier $\lambda_1^*$, 
which is just that for the 1D case. However, the second Floquet 
multiplier $\lambda_2^*$ affected by coupling depends on the  
fixed maps; $\lambda_2^*=0$ $(1)$ for the $I$ $(E)$ map.

In order to confirm the above renormalization results, we also 
study the intermittency for the linearly-coupled case. The critical
set (set of critical points) for the intermittency consists of 
critical line segments. It is found that the $I$ map with no
relevant CE's governs the critical behavior at interior points of 
each critical line segment, while the $E$ map with one relevant CE
$\delta_2$ $(=2)$ governs the critical behavior at both ends.

We choose $f(x)=1-a x^2$ as the uncoupled 1D map in Eq.~(\ref{eq:TCM})
and consider a linear coupling case $g(x,y)=c(y-x)$. Here $c$ is a 
coupling parameter. Three critical line segments are found on a 
synchronous saddle-node bifurcation line $a=a_c$ $(=1.75$, above which 
a pair of synchronous orbits with period 3 appears. The critical 
behaviors near the three critical line segment are the same.

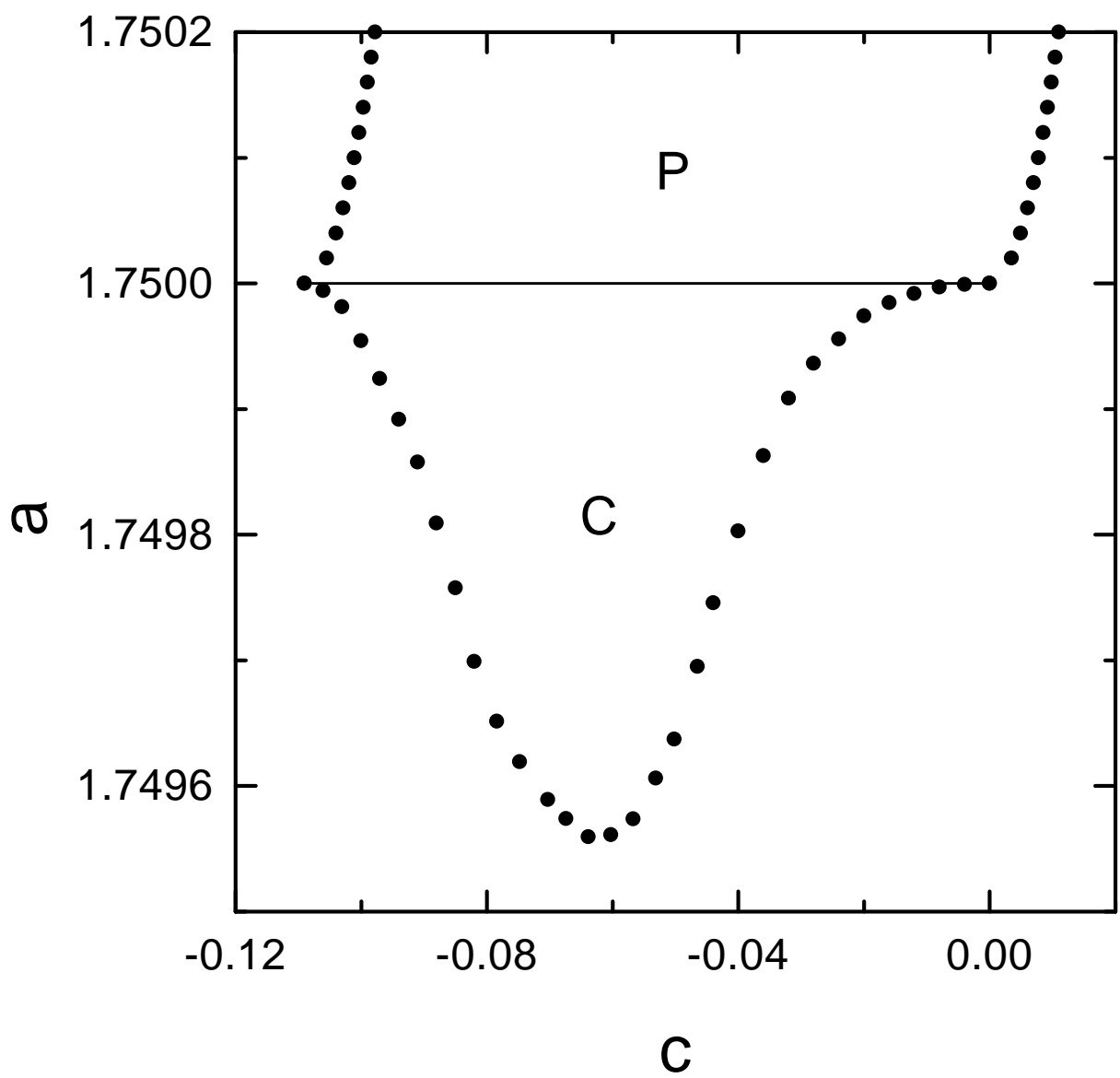
\begin{figure}[tbp]
\caption{ Phase diagram of the two-coupled 1D map (1) with 
$f(x)=1-ax^2$ and $g(x,y)=c(y-x)$. Here solid circles denote the data 
points on the $\sigma_\bot=0$ curve. The region enclosed by the 
$\sigma_\bot =0$ curve is divided into two parts denoted by P and C. A 
synchronous period-3 (chaotic) attractor with $\sigma_\| <0$ 
$(\sigma_\| >0)$ exists in the subregion P (C). The boundary curve 
denoted by a solid line between the P and C regions is just a critical
line segment.}
\label{PD}
\end{figure}

As an example, consider a critical line segment including the 
zero-coupling point $c=0$ as one end point. Figure \ref{PD} shows 
a phase diagram near this critical line segment denoted by a solid 
line. This diagram is obtained from the calculation of two Lyapunov 
exponents. In case of a synchronous orbit, its Lyapunov exponents are 
given by 
\begin{equation}
\sigma_\| (a) = {\lim_{m \rightarrow \infty}}\, {\frac{1 }{m}}
\sum_{n=0}^{m-1} \ln|f^{\prime}(x_n)|,\;\; \sigma_\bot (a,c) = 
{\lim_{m \rightarrow \infty}}\, {\frac{1 }{m}} \sum_{n=0}^{m-1}
\ln|f^{\prime}(x_n)-2c|.  
\label{eq:LE}
\end{equation}
Here $\sigma_\|$ $(\sigma_\bot)$ denotes the mean exponential rate
of divergence of nearby orbits along (across) the synchronization 
line $y=x$. Hereafter, $\sigma_\|$ and $\sigma_\bot$ will be referred 
to as tangential and transversal Lyapunov exponents, respectively. 
Note also that $\sigma_\|$ is just the Lyapunov exponent for the 
1D case, and the coupling affects only $\sigma_\bot$.

The data points on the $\sigma_\bot=0$ curve are denoted by solid 
circles in Fig.~\ref{PD}. A synchronous orbit on the synchronization 
line $y=x$ becomes a synchronous attractor with $\sigma_\bot <0 $ 
inside the $\sigma_\bot=0$ curve. The type of this synchronization 
attractor is determined according to the sign of $\sigma_\|$. A 
synchronous period-3 orbit with $\sigma_\| < 0$ becomes a synchronous
periodic attractor above the critical line segment, while there exists 
a synchronous chaotic attractor with $\sigma_\| >0$ below the critical 
line segment. These periodic and chaotic regions are denoted by P and 
C in the diagram, respectively. There exists a synchronous period-3 
attractor with $\sigma_\| =0$ on the critical line segment between 
these two regions.

The motion on the synchronous chaotic attractor in the region C just 
below the critical line segment is characterized by the occurrence of 
intermittent alternations between regular behavior and chaotic 
behavior on the synchronization line. This is just the intermittency 
occurring in the uncoupled 1D map, because the motion on the 
synchronization line is the same as that for the uncoupled 1D case. 
Thus, a transition from a regular behavior to an intermittent
chaotic behavior, which is essentially the same as that for the 1D 
case, occurs near the critical line segment joining two end points 
$c_l = -0.109045 \cdots$ and $c_r=0$ on the synchronous saddle-node
bifurcation line $a=a_c(=1.75)$.

Consider a ``1D-like'' intermittent transition to chaos near an 
interior point with $c_l < c < c_r$ of the critical line segment.  
We fix the value of $c$ at some interior point and vary the control 
parameter $\epsilon$ $(\equiv a_c -a)$. For $\epsilon <0$, there 
exists a synchronous period-3 attractor on the synchronization
line. However, as $\epsilon$ is increased from zero, an intermittent 
synchronous chaotic attractor appears. Like the 1D case \cite{HH}, the 
scaling relations of the mean duration $\bar l$ of regular behavior 
and the tangential Lyapunov exponent $\sigma_\|$ for an intermittent 
chaotic orbit on the synchronization line are obtained from the 
leading relavant eigenvalue $\delta_1$ $(=4)$ of the $I$ map, as will 
be seen below.

We first note that the $I$ map is essentially a 1D map with zero 
Jacobian determinant (see Table \ref{FP}). Since there exists no 
relevant CE's associated with coupling perturbation, it has only 
relevant eigenvalues $\delta_1$ and $\delta_1^{\prime}$ like the 1D 
case. The $I$ map is therefore associated with the critical behavior 
at interior points of the critical line segments.

A map with non-zero $\epsilon$ near a critical interior point is 
transformed to a new map of the same form, but with a new parameter 
$\epsilon ^{\prime}$ under a renormalization transformation 
${\cal R}$. Here the control parameter scales as 
\begin{equation}
\epsilon ^{\prime}= \delta_1 \, \epsilon \,=\, 2^2 \epsilon.
\end{equation}
Then the mean duration $\bar l$ and the tangential Lyapunov exponent 
$\sigma_\|$ satisfy the homogeneity relations, 
\begin{equation}
{\bar l} (\epsilon ^{\prime}) = {\frac{1 }{2}} {\bar l} (\epsilon),
\;\;\; {\sigma_\|} (\epsilon ^{\prime}) = 2 {\sigma_\|}(\epsilon),
\end{equation}
which lead to the scaling relations, 
\begin{equation}
{\bar l} (\epsilon) \sim \epsilon ^{- \mu},\;\;\; {\sigma_\|}
(\epsilon) \sim \epsilon ^{\mu},  
\label{eq:SR}
\end{equation}
with exponent 
\begin{equation}
\mu = {\log 2} / \log {\delta_1} = 0.5.
\end{equation}

The above 1D-like intermittent transition to chaos ends at both ends 
of the critical line segment. We fix the value of the control 
parameter $a=a_c$ $(=1.75)$ and study the critical behavior near both 
ends $c_l$ and $c_r$ by varying the coupling parameter $c$. Inside 
the critical line segment $(c_l < c < c_r)$, a synchronous period-3 
attractor with $\sigma_\bot <0$ exists on the synchronization line, 
and hence the coupling tends to synchronize the interacting 
systems. However, as the coupling parameter $c$ passes through both 
ends, the transversal Lyapunov exponent $\sigma_\bot$ of the
synchronous periodic orbit grows continuously from zero, and hence the
coupling leads to desynchronization of the interacting systems. The
synchronous orbit of period 3 is therefore no longer an attractor 
outside the critical line segments, and a new asynchronous attractor 
appears.

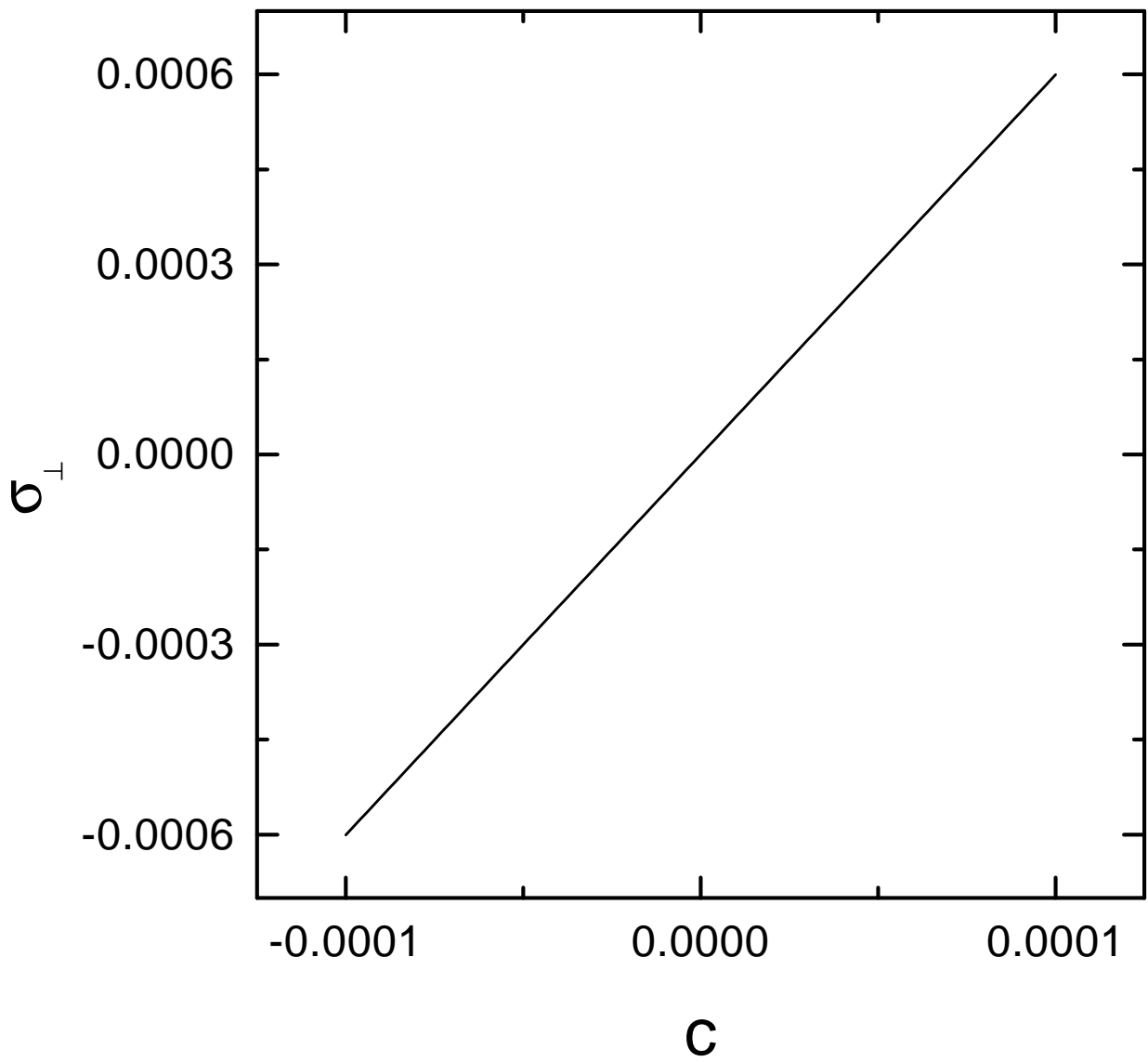
\begin{figure}[tbp]
\caption{ Plot of the transversal Lyapunov exponent $\sigma_\bot$ of 
 a synchronous period-3 orbit versus $c$ for $a=a_c$ $(=1.75)$. }
\label{CLexp}
\end{figure}

The critical behaviors near both ends are the same. As an example, 
consider the case of the zero-coupling point $c_r=0$. Figure 
\ref{CLexp} shows the plot of $\sigma_\bot$ versus $c$ for $a=a_c$. 
Note that $\sigma_\bot$ increases linearly with respect to $c$. Hence 
a transition from a synchronous to an asynchronous state occurs at the 
zero-coupling end point.

The scaling relation of $\sigma_\bot (c)$ for $a=a_c$ is obtained from 
the relevant CE $\delta_2$ $(=2)$ of the $E$ map as follows. Consider 
a map with non-zero $c$ near the zero-coupling point. It is then 
transformed to a map of the same form, but with a renormalized
parameter $c^{\prime}$ under a renormalization transformation 
${\cal R}$. Here the coupling parameter obeys a scaling law, 
\begin{equation}
c^{\prime}= \delta_2 c = 2 c.
\end{equation}
Then the transversal Lyapunov exponent $\sigma_\bot$ satisfies the 
homogeneity relation, 
\begin{equation}
\sigma_\bot (c^{\prime}) = 2 \sigma_\bot (c).
\end{equation}
This leads to the scaling relation, 
\begin{equation}
\sigma_\bot (c) \sim c^{\nu},
\end{equation}
with exponent 
\begin{equation}
\nu = {\log 2} / {\log \delta_2} =1.
\end{equation}

Like the case of the $I$ map, the scaling behavior of 
$\sigma_\| (\epsilon)$ for $c=c_l$ or $c_r$ is obtained from the 
relevant eigenvalue $\delta_1$ $(=4)$ of the $E$ map, and hence it 
also satisfies the scaling relation (\ref{eq:SR}). The critical 
behaviors of both exponents $\sigma_\|$ and $\sigma_\bot$ near an end 
point are thus determined from two relevant eigenvalues $\delta_1$ 
and $\delta_2$ of the $E$ map. An extended version of this work 
including the results of a renormalization analysis without 
truncation, the results for the many-coupled cases and so on will be 
given elsewhere \cite{Kim5}

\acknowledgments
This work was supported by the the Korea Research Foundation under 
Project No. 1997-001-D00099.

\mediumtext
\begin{table}[tbp]
\caption{ Fixed point ${\bf P}^*$ of the renormalization 
 transformation ${\cal R}$ and the rescaling factor $\alpha$. {} }
\label{FP}
\begin{tabular}{cccccccc}
fixed map & $\alpha$ & $A^*$ & $B^*$ & $C^*$ & $D^*$ & $E^*$ & 
$F^*$ \\ 
\tableline $I$ map & 2 & 0 & 1 & arbitrary & 0 & 0 & 0 \\ 
$E$ map & 2 & 0 & 1 & arbitrary & 1 & arbitrary & arbitrary   
\end{tabular}
\end{table}

\mediumtext
\begin{table}[tbp]
\caption{ Some eigenvalues $\delta_1, \delta^{\prime}_1, \delta_2,$ 
and $\delta^{\prime}_2$ of a fixed map $T^*$ of the renormalization 
operator are shown.}
\label{EV}
\begin{tabular}{ccccc}
fixed map & $\delta_1$ & $\delta^{\prime}_1$ & $\delta_2$ & 
$\delta^{\prime}_2$ \\ 
\tableline $I$ map & 4 & 2 & nonexistent & nonexistent \\ 
$E$ map & 4 & 2 & 2 & 1  
\end{tabular}
\end{table}

\narrowtext

\end{document}